\documentclass[12pt,a4paper]{article}
\usepackage{ifthen}
\usepackage{graphicx}

\usepackage{subfigure}
\usepackage{cite}
\usepackage{doublespace}
\usepackage{Vmargin}

\newcommand{\capa}{}\newcommand{\capb}{}
\newcommand{\pairoffigures}[4][tree]{%
  \renewcommand{\capa}{}\renewcommand{\capb}{}
  \ifthenelse{\equal{#1}{tree}}{\renewcommand{\capa}{probability tree}}{}%
  \ifthenelse{\equal{#1}{ds}}{\renewcommand{\capa}{consistency statistics}}{}%
  \ifthenelse{\equal{#1}{pjt}}{\renewcommand{\capa}{projection times}}{}%
  \ifthenelse{\equal{#2}{tree}}{\renewcommand{\capb}{probability tree}}{}%
  \ifthenelse{\equal{#2}{ds}}{\renewcommand{\capb}{consistency  statistics}}{}%
  \ifthenelse{\equal{#2}{pjt}}{\renewcommand{\capb}{projection times}}{}%
  \begin{figure}[ht]
    \begin{center}
      \begin{tabular}{cc}
        \subfigure[\capa]{\includegraphics{#3.#1.eps}} &
        \subfigure[\capb]{\includegraphics{#3.#2.eps}}
      \end{tabular}
    \end{center}
    \caption{#4}
    \label{fig:#3}
  \end{figure}}

\newcommand{\onefigure}[3][tree]{%
  \renewcommand{\capa}{}
  \ifthenelse{\equal{#1}{tree}}{\renewcommand{\capa}{probability tree}}{}%
  \ifthenelse{\equal{#1}{ds}}{\renewcommand{\capa}{consistency statistics}}{}%
  \ifthenelse{\equal{#1}{pjt}}{\renewcommand{\capa}{projection times}}{}%
  \begin{figure}[ht]
    \centering
    \includegraphics{#2.#1.eps}
    {\small \capa}
    \caption{#3}
    \label{fig:#2}
  \end{figure}}
\begin{document}

\title{Random Hamiltonian Models and Quantum Prediction Algorithms}

\author{Jim McElwaine\thanks{E-mail: jnm11@damtp.cambridge.ac.uk}}


\date{18th Decmeber 1998}
\maketitle
\begin{abstract}
  This paper describes an algorithm for selecting a consistent set
  within the consistent histories approach to quantum mechanics and
  investigates its properties. The algorithm select from among the
  consistent sets formed by projections defined by the Schmidt
  decomposition by making projections at the earliest possible time.
  The algorithm unconditionally predicts the possible events in closed
  quantum systems and ascribes probabilities to these events.  A
  simple random Hamiltonian model is described and the results of
  applying the algorithm to this model using computer programs are
  discussed and compared with approximate analytic calculations.
\end{abstract}

\section{Introduction}

It is hard to find an entirely satisfactory interpretation of the
quantum theory of closed systems, since quantum theory does not
distinguish physically interesting time-ordered sequences of
operators. In this paper, we consider one particular line of attack on
this problem: the attempt to select consistent sets by using the
Schmidt decomposition together with criteria intrinsic to the
consistent histories formalism. For a discussion of why we believe
consistent histories to be incomplete without a set selection
algorithm see~\cite{Dowker:Kent:properties,Dowker:Kent:approach} and
for other ideas for set selection algorithms
see~\cite{McElwaine:Kent,McElwaine:2,gmhstrong,Isham:Linden:information}.
This issue is controversial: others believe that the consistent
histories approach is complete in
itself~\cite{omnesbook,griffithschqr,CEPI:GMH}.

\subsection{Consistent histories formalism}\label{ssec:CHformalism} 

We use a version of the consistent histories formalism in which the
initial conditions are defined by a pure state, the histories are
branch-dependent and consistency is defined by Gell-Mann and Hartle's
medium consistency criterion eq.~(\ref{mediumcon}).  We restrict
ourselves to closed quantum systems with a Hilbert space in which we
fix a split ${\cal H} = {\cal H}_1 \otimes {\cal H}_2$; we write $\dim
({\cal H}_j ) = d_j$ and we suppose that $d_1 \leq d_2 < \infty$.  The
model described in sec.~\ref{sec:random:intro} has a natural choice for
the split. Other possibilities are discussed in~\cite{McElwaine:Kent}.

Let $|\psi\rangle$ be the initial state of a quantum system.  A
\emph{branch-dependent set of histories} is a set of products of
projection operators indexed by the variables $\alpha = \{ \alpha_n ,
\alpha_{n-1} , \ldots , \alpha_1 \}$ and corresponding time
coordinates $\{ t_n , \ldots , t_1 \}$, where the ranges of the
$\alpha_k$ and the projections they define depend on the values of
$\alpha_{k-1} , \ldots , \alpha_1 $, and the histories take the form:
\begin{equation} \label{histories}
  C_{\alpha} = P_{\alpha_n}^n (t_n ; \alpha_{n-1} , \ldots , \alpha_1)
  P_{\alpha_{n-1}}^{n-1} (t_{n-1} ; \alpha_{n-2} , \ldots , \alpha_1)
  \ldots P_{\alpha_1}^1 ( t_1 )\,.
\end{equation}
Here, for fixed values of $\alpha_{k-1} , \ldots , \alpha_1$, the
$P^k_{\alpha_k} (t_k ; \alpha_{k-1} , \ldots , \alpha_1 )$ define a
projective decomposition of the identity indexed by $\alpha_k$, so
that $\sum_{\alpha_k} P^k_{\alpha_k} (t_k ; \alpha_{k-1} , \ldots ,
\alpha_1 ) = 1 $ and
\begin{equation} \label{decomp}
P^k_{\alpha_k} (t_k ; \alpha_{k-1} , \ldots , \alpha_1 )
P^k_{\alpha'_k} (t_k ; \alpha_{k-1} , \ldots , \alpha_1 ) =
\delta_{\alpha_k \alpha'_k } P^k_{\alpha_k} (t_k ; \alpha_{k-1} ,
\ldots , \alpha_1 )\,.
\end{equation}
Here and later, though we use the compact notation $\alpha$ to refer
to a history, we intend the individual projection operators and their
associated times to define the history.

We use the consistency criterion\footnote{For a discussion of other
  consistency criteria see, for example,
  refs.~\cite{Kent:implications,Goldstein:Page,%
    Dowker:Halliwell,McElwaine:1}.}
\begin{equation}
  D_{\alpha\beta} = 0, \quad \forall \alpha \neq \beta,
  \label{mediumcon}
\end{equation}
which Gell-Mann and Hartle call \emph{medium consistency}, where
$D_{\alpha\beta}$ is the \emph{decoherence matrix}
\begin{equation}
  D_{\alpha\beta} = \mbox{Tr}\, (C_\alpha \rho C_\beta^\dagger)\,.
  \label{dmdef}
\end{equation}
Probabilities for consistent histories are defined by the formula
\begin{equation}\label{probdef}
  p(\alpha) = D_{\alpha\alpha}.
\end{equation}

With respect to the ${\cal H} = {\cal H}_1 \otimes {\cal H}_2$
splitting of the Hilbert space, the \emph{Schmidt decomposition} of
$|\psi (t) \rangle$ is an expression of the form
\begin{equation} \label{schmidteqn}
  |\psi (t) \rangle = \sum_{i=1}^{d_1} \, [p_i(t)]^{1/2} \, | w_i
  (t)\rangle_1 \otimes |w_i (t)\rangle_2 \, ,
\end{equation}
where the \emph{Schmidt states} $\{ |w_i\rangle_1 \}$ and $\{
|w_i\rangle_2\}$ form, respectively, an orthonormal basis of ${\cal
H}_1$ and part of an orthonormal basis of ${\cal H}_2$, the functions
$p_i (t)$ are real and positive, and we take the positive square root.
For fixed time $t$, any decomposition of the form
eq.~(\ref{schmidteqn}) then has the same list of probability weights
$\{ p_i (t) \}$, and the decomposition~(\ref{schmidteqn}) is unique if
these weights are all different. These probability weights are the
eigenvalues of the reduced density matrix.

The idea motivating this paper is that the combination of the ideas of
the consistent histories formalism and the Schmidt decomposition might
allow us to define a mathematically precise and physically interesting
description of the quantum theory of a closed system. We consider
constructing histories from the projection operators\footnote{There
are other ways of constructing projections from the Schmidt
decomposition~\cite{McElwaine:Kent}, though for the model considered
in this paper the choices are equivalent.}
\begin{equation} \label{schmidtprojs1}
  \begin{array}{lll}
  P_i (t) = | w_i (t) \rangle_1 \langle w_i (t) |_1 \otimes I_2
  &\mbox{and}& \overline P = I_1 \otimes I_2 - \sum_i P_i (t)\, ,
 \end{array}
\end{equation}
which we refer to as \emph{Schmidt projections}. If $\mbox{dim}{\cal
  H}_1 = \mbox{dim}{\cal H}_2$ the complementary projection $\overline
  P$ is zero. In developing the ideas of this paper, we were
  influenced in particular by Albrecht's
  investigations~\cite{Albrecht:decoherence,Albrecht:collapsing} of
  the behaviour of the Schmidt decomposition in random Hamiltonian
  interaction models and the description of these models by consistent
  histories.

\section{A Random Hamiltonian Model}
\label
{sec:random:intro}
Consider a simple quantum system consisting of a finite Hilbert space
$\mathcal{H} = \mathcal{H}_1 \otimes \mathcal{H}_2$
($\mbox{dim}\mathcal{H}_i = d_i$), a pure initial state
$|\psi(0)\rangle$ and a Hamiltonian drawn from the GUE (Gaussian
Unitary Ensemble), which is defined by
\begin{equation}
  \label{GUEa}
  P(H) = N^{-1} \exp\{-\mbox{Tr}[(\lambda H + \mu)^2]\},
\end{equation}
where $N$ is a normalisation constant.

The GUE is the unique ensemble of Hermitian matrices invariant under
$U(d)$ with independently distributed matrix elements, where $d = \mbox{dim}
{\cal H} =d_1d_2$.  The GUE is also the unique ensemble with maximum entropy,
$-\int dH\, P(H) \log P(H)$, subject to $E[\mbox{Tr}(H)] = \mu$ and
$E[\mbox{Tr}(H^2)] = \lambda$.  The book by Mehta~\cite{Mehta}
contains a short proof of this as well as further analysis of the GUE
and related ensembles. All the results concerning the GUE in this
thesis can be found in this book or in the appendix.

This model is not meant to represent any particular physical system,
though Hamiltonians of this from are used in models of nuclear
structure and have often been studied in their own right
(see~\cite{Mehta,Simons:Altshuler} and references therein), and a
large class of other ensembles approximate the GUE in the large $d$
limit.

Because $H$ is drawn from a distribution invariant under $U(d)$ there
is no preferred basis, no distinction between system and environment
degrees of freedom and no time asymmetry. In other words the model is
chosen so that there is no obvious consistent set: we do not already
know what the answer should be.  Moreover it does not single out a
pointer basis that one might associate with classical states, so that
the Copenhagen interpretation cannot make any predictions about a
model like this in the $t \to \infty$ limit.  If an algorithm works
for this model, when there are no special symmetries, it should work
for a wide variety of models.  The question whether a pointer basis
can arise dynamically using Schmidt states was addressed by Albrecht
in~\cite{Albrecht:decoherence,Albrecht:collapsing}, but no general
prescription emerged from his study. Albrecht also studied the
relationship between Schmidt states and consistent histories, and his
studies suggested that the relationship was complicated.  

The model
considered here generalises Albrecht's model: the Hamiltonian for the
entire Hilbert space is chosen from a random ensemble. Albrecht also
used a different distribution, but as we explain below the GUE seems
more natural, though this it probably makes little difference.
 
Without loss of generality, we take $\mu=0$ and
$\lambda=1/2$. With this choice and using the Hermiticity property of
$H$ eq.~(\ref{GUEa}) becomes
\begin{equation}
  \label{GUEb}
   P(H) = \pi^{-n^2/2} 2^{-n/2} N^{-1} \prod_{i<j} e^{-|H_{ij}|^2}
   \prod_{i} e^{-|H_{ii}|^2/2}.
\end{equation}
Therefore the diagonal elements are independently distributed, \emph{real}
normal random variables with mean $0$ and variance $1$ and the
off-diagonal elements are independently distributed, \emph{complex}
normal random variables with mean $0$ and variance $1$.

Since the Hamiltonian is invariant under $U(d)$ the only significant
degrees of freedom in the choice of initial state are the initial
Schmidt eigenvalues (the eigenvalues of the initial reduced density
matrix.)  The usual choice in an experimental situation is an initial
state of the form $|\psi\rangle = |u\rangle_1 \otimes |v\rangle_2$
which corresponds to a pure initial density matrix. A more general
choice in the spirit of the model is to draw the initial state from
the $U(d)$ invariant distribution subject to fixed rank $n$. This is
equivalent to choosing the first $n$ eigenvalues to be components of a
random unit vector in $R^n$ and the remaining $d_1-n$ components to be
zero.

\section{Analysis}\label{sec:rand:analysis}
The calculations in this section are an attempt to gain insight into
the expected properties of prediction algorithms applied to the random
model. These calculations rest on a large number of assumptions and
are at best approximations, but the conclusions are borne out by
numerical simulations and the calculations do provide a rough feel for
the results that different algorithms can be expected to produce.  In
particular they suggest that there are only narrow ranges of values
for the approximate consistency parameter which are likely to produce
physically plausible sets of histories.  These calculations may also
be applicable to other models since this model makes so few
assumptions and the interaction is completely general.

In a random model there is no reason to expect exactly consistent sets
of histories formed from Schmidt projections to exist, so only
parameterised approximate consistency 
criteria such as the frequently used
criterion~\cite{GM:Hartle:classical,Omnes:5}
\begin{equation}
\label{badcriterion}
  |D_{\alpha\beta}| \leq \epsilon(\delta), \quad \forall \alpha \neq
\beta
\end{equation} 
or the DHC (Dowker-Halliwell Criterion)~\cite{Dowker:Halliwell,McElwaine:1}
 \begin{eqnarray}\label{wic:DHC}
  \left|\mbox{Re}\,(D_{\alpha\beta})\right| & \leq & \epsilon \,
    (D_{\alpha\alpha}D_{\beta\beta})^{1/2}, \quad \forall\,
    \alpha\neq\beta,
\end{eqnarray}
or
\begin{eqnarray}\label{DHC}
  |D_{\alpha\beta}| & \leq & \epsilon \,
    (D_{\alpha\alpha}D_{\beta\beta})^{1/2}, \quad \forall\,
    \alpha\neq\beta.
\end{eqnarray}
 are considered in this paper and  $\epsilon$ will always be the consistency
parameter in these equations.  We shall only discuss medium consistency
criteria: the results for weak consistency are qualitatively the same.

Approximate consistency criteria were analysed further in ref.
\cite{McElwaine:1}.  As refs. \cite{Dowker:Halliwell,McElwaine:1}
explain, the DHC has natural physical properties and is well adapted
for mathematical analyses of consistency.  We adopt it here, and refer
to the largest term,
\begin{equation} \label{dhp} 
\mbox{max} \{\,
  |D_{\alpha\beta}|(D_{\alpha\alpha}D_{\beta\beta})^{-1/2} \, : \,
  \alpha , \beta \in S \, , \alpha \neq \beta \, , \mbox{~and~}
  D_{\alpha\alpha} , D_{\beta\beta} \neq 0 \,\} \, ,
\end{equation} 
of a (possibly incomplete) set of histories $S$ as the
Dowker-Halliwell parameter, or DHP.

If an absolute approximate consistency criterion is being used there
are strong theoretical reasons for imposing a parameterised
non-triviality criterion~\cite{McElwaine:1}. However,
if the approximate DHC is being used one is not needed, though it is
convenient to introduce one for computational reasons. The
non-triviality parameter (which we shall always write as $\delta$ in
this chapter) can be taken very small if the DHC is used and is not
expected to influence the results --- except possibly for the first
projections --- and the numerical simulations show that this is indeed
the case.  We shall refer to histories with probability less than or
equal to $\delta$ (relative or absolute) as \emph{trivial} histories
and a projection that gives rise to a \emph{trivial} history as a
\emph{trivial} projection. There are no absolute reasons for rejecting
set of histories containing trivial histories --- if $\delta$ is
sufficiently small and there are not too many they are physically
irrelevant --- though obviously sets are preferable if all the
histories are non-trivial. However, an algorithm must produce results
that are approximately the same for a range of parameter values if it
is to make useful predictions, and trivial histories will almost
certainly vary as $\delta$ is changed.  If the DHC is used,
generically all the later projections will also change, since trivial
histories can significantly influence the consistency of later
projections.  If an absolute consistency criterion is used trivial
projections are more likely to be consistent than non-trivial
projections so for many values of the parameters only trivial will be
projections are made.

\subsection{Repeated projections and relative consistency}
Consider a history $\alpha$ extended by the projective decomposition
$\{P, \overline P\}$ and the further extension of history
$P|\alpha\rangle$ by $\{P(t), \overline P(t)\}$. This was discussed
for the DHC in refs.~\cite{McElwaine:1,McElwaine:Kent} and the DHP for
this case was shown to be
\begin{equation}\label{repeatedDHCa}
  \frac{|\langle\alpha| \overline{P} \dot{P} |\alpha\rangle| }{\|
    \overline{P} |\alpha\rangle \| \, \| \overline{P} \dot{P}
    |\alpha\rangle \|}.
\end{equation}
The reprojection will occur unless $\epsilon$, the approximate
consistency parameter, is smaller than (\ref{repeatedDHCa}).  It is
easy to show that the time evolution of Heisenberg picture Schmidt
projections is
\begin{equation}\label{anl:pevol}
  \dot P = i[H-B \otimes I,P],
\end{equation}
where $H$ is the Hamiltonian,
\begin{equation}
  B = \sum_{k \neq m} \frac{Q_k \rho_r Q_m}{p_m-p_k} \, ,
\end{equation}
$Q_k$ are projection operators (in $H_1$) on to the Schmidt
eigenspaces, $p_k$ their respective (distinct) eigenvalues and
$\dot \rho_r$ the derivative of the reduced density matrix. 

In analysing (\ref{repeatedDHCa}) and similar expressions we
make the following assumptions. First that $|\alpha\rangle$ is
uncorrelated with the Schmidt states --- this generally is a good
approximation when there are a large number of histories.  Second
that $B \otimes I$ is an operator drawn from the GUE with unspecified
variance independent of the other variables --- in some situations
this assumption is exact but it generically is not.

Let $G = H-B \otimes I$, an element of the GUE with variance $\sigma$,
then using eq.~(\ref{anl:pevol}) (\ref{repeatedDHCa}) is
\begin{equation}\label{repeatedDHCb}
  \frac{|\langle\alpha| \overline P G P |\alpha\rangle| }{\|
    \overline P |\alpha\rangle \| \, \| \overline P G P |\alpha\rangle
    \|}.
\end{equation}
Because $G$ is drawn from a distribution invariant under $U(d)$ and is
independent of $P|\alpha\rangle$ and $\overline
P|\alpha\rangle$, (\ref{repeatedDHCb}) can be simplified by
choosing a basis in which $P|\alpha\rangle/\|P|\alpha\rangle\| =
(1,\ldots,0)$ and $\overline P|\alpha\rangle/\|\overline
P|\alpha\rangle\| = (0,1,\ldots,0)$. (\ref{repeatedDHCb}) becomes
\begin{equation}\label{repeatedDHCc}
  \frac{|Z_{1}|}{[\sum_{r \geq k \geq 1} |Z_k|^2]^{1/2}}\,,
\end{equation}
where $r =\mbox{rank}(\overline P)$ and $Z_k =G_{1(k+1)}$. Since
$\{G_{ij},i < j\}$ is a set of independent, complex, normal random
variables, (\ref{repeatedDHCc}) is the square root of a $B(1,r-1)$
random variable\footnote{$B(p,q)$ :- a beta random variable with
  parameters $p$ and $q$. This has a density function $\propto
  t^{p-1}(1-t)^{q-1}$. A B(1,r-1) random variable has the same
  distribution as that of the inner-product squared between two
  independent unit vectors in $C^r$.}.  

Suppose we choose $\epsilon$ so that reprojections will occur with
some small probability $q$ --- note that only choosing $\epsilon = 0$
will definitely prevent all repeated projections.  The probability of
(\ref{repeatedDHCc}) being less than $\epsilon$ is
\begin{equation}
  1 - (1-\epsilon^2)^{r-1}.
\end{equation}
Therefore if
\begin{eqnarray}\nonumber
  \epsilon &\approx& [1-(1-q)^{1/(r-1)}]^{1/2}
  = \left[\frac{-\log(1-q)}{r}\right]^{1/2}
  + O \left[\frac{\log(1-q)}{r}\right]^{3/2} 
  \\ &&\leq d^{-1/2} \sqrt{-\log(1-q)}\,,\label{repeatedDHCd} 
\end{eqnarray}
a reprojection will occur with probability $\approx q$.

However, it is shown in~\cite{McElwaine:Kent} that the DHC
cannot prevent trivial reprojections on the initial state if the
initial density matrix has less than full rank. If the initial density
matrix has rank one then the first projection will always be made with
probability $\delta$. A non-triviality criterion can then work in
conjunction with the DHC to prevent further trivial extensions.
Suppose either that the initial density matrix has rank greater than
$1$ and $P_n$ and $P_m$ are two projections onto the non-zero
eigenspaces, or assume that the rank is one and $P_n$ is a projection
onto the initial state and $P_m$ is a projection making a history of
probability $\delta$. In either case, let $P_k$ be projection onto the
null space.  To prevent the trivial projection $P_k$ being made the
parameters $\delta$ and $\epsilon$ must be chosen to 
satisfy~\cite{McElwaine:Kent}
\begin{equation}\label{edinequalityr}
  \sqrt{\delta} |\langle \psi | P_{m} \dot P_{k}^2 P_{n} | \psi
  \rangle| > \epsilon \| P_{m} | \psi \rangle \|\, \| \dot P_{k} P_{n}
  | \psi \rangle \|^2.
\end{equation}
Though the probability distribution for this is complicated, the
approximate relation between $\delta$ and $\epsilon$ can be estimated
by squaring both sides of eq.~(\ref{edinequalityr}) and then taking
the expectation. Note that treating $G$ as an element of the GUE is
exact in this case as the terms involving $B \otimes I$ are
identically zero. Using results from eq.~(\ref{expectgue}),
eq.~(\ref{edinequalityr}) becomes
\begin{equation}\label{preventinitial:rel}
  \delta > \epsilon^2 (r+1) \|P_n|\psi\rangle\|^2\,,
\end{equation}
where $r$ is the rank of $P_k$. By assumption $\|P_n|\psi\rangle\|$ is
order one and $r<d$, so if $\delta > d \epsilon^2$ initial
reprojections will not occur. The results are the same for a relative
non-triviality criterion since instead of
eq.~(\ref{preventinitial:rel}) we have $ \delta > \epsilon^2 (r+1)$.

\subsection{Repeated projections and absolute consistency}
An algorithm using an absolute parameterised consistency criterion
will make nothing but trivial projections unless a parameterised
non-triviality criterion is also used, so only algorithms with a
non-triviality criterion are considered.

Let $t_\epsilon$ denote the latest time that the reprojection is
approximately consistent and $t_\delta$ the earliest time at which the
extension is absolutely nontrivial. We see from
refs.~\cite{McElwaine:1,McElwaine:Kent} that, to
lowest order in $t$,
\begin{eqnarray}
  t_{\delta} &=& \sqrt{\delta} \|\dot P P | \alpha \rangle \|^{-1} \\
  t_{\epsilon} &=& \epsilon |\langle \alpha | \overline P \dot P P |
  \alpha \rangle|^{-1}\,.
\end{eqnarray}
$t_{\delta} > t_{\epsilon}$ implies
\begin{equation}\label{absoluteta}
  \sqrt{\delta} |\langle \alpha | \overline P \dot P P | \alpha
  \rangle| > \epsilon \| \dot P P |\alpha\rangle \|.
\end{equation}
Again we choose $\epsilon$ so that reprojections occur with
probability q and assume that $\| P |\alpha\rangle \|$ and $\|
\overline P |\alpha\rangle \|$ are order one, so that
eq.~(\ref{absoluteta}) can be written
\begin{equation}\label{absolutetb}
  \frac{|Z_{1}|}{[\sum_{r \geq k \geq 1} |Z_k|^2]^{1/2}} >
  \epsilon/\sqrt{\delta}.
\end{equation}
The l.h.s.\ is the same random variable as in eq.~(\ref{repeatedDHCc})
so $\delta$ and $\epsilon$ must be chosen so that 
\begin{equation}\label{prevrepabsa}
  \epsilon/\sqrt \delta = d^{-1/2} \sqrt{-\log(1-q)}\,.
\end{equation}
The assumption that $\| P |\alpha\rangle \|$ and $\| \overline P
|\alpha\rangle \|$ are order one will obviously not always be valid.
As more projections are made the probabilities of the histories will
decrease. When both probabilities  are
$\delta$ eq.~(\ref{absoluteta}) is
\begin{equation}\label{absolutetc}
  \frac{|Z_{1}|}{[\sum_{r \geq k \geq 1} |Z_k|^2]^{1/2}} >
  \epsilon/\delta,
\end{equation}
so $\epsilon/\delta = d^{-1/2} \sqrt{-\log(1-q)}$. If reprojections of
smaller probability histories are to be prevented this choice of
parameters is clearly more appropriate than eq.~(\ref{prevrepabsa}).

This analysis has picked a very conservative upper bound for
$\epsilon$ to prevent repeated projections, since decoherence matrix
terms with the other histories will tend to reduce the likelihood of
repeated projections, and thus allow larger values of $\epsilon$ to be
used. A more detailed analysis suggests that for relative and absolute
consistency $r$ can be treated as much smaller than $d$ so that
choosing $\delta$ a small factor larger than $\epsilon^2$ or
$\epsilon$ respectively, are sufficient conditions.

\subsection{Projections in the long time limit}\label{subsec:longtime}
The previous subsection has shown how $\epsilon$ and $\delta$ affect
the probability of repeated projections: this subsection calculates
how they affect the probability of projections as $t \to \infty$. In
infinite dimensional systems, off-diagonal terms of the decoherence
matrix for quasiclassical projections often tend to zero as $t$
increases~\cite{Zurek:superselection}. In the limit $d \to \infty$ one
would also expect this for Schmidt projections in this model ---
though the limit only exists for initial density matrices of finite
rank. 

Consider the DHP for a Schmidt projection extending history $\alpha$
from the set of normalised exactly consistent histories
$\{|\alpha\rangle,|\beta_i\rangle, i = 1,\ldots, k\}$ as $t\to\infty$.
For $t\to \infty$ and for large $k$ the Schmidt states are
approximately uncorrelated with the histories. The DHP for an
extension $\{P, \overline P\}$ of history $\alpha$ is
\begin{equation}\label{rellongtime}
 \max \left \{
 \frac{|\langle\beta_i|P|\alpha\rangle|}{\|P|\alpha\rangle\|},
 \frac{|\langle\beta_i|\overline P|\alpha\rangle|}{%
 \|\overline P|\alpha\rangle\|}, i = 1,\ldots,k \right \}. 
\end{equation}
Since $\langle\beta_i|\alpha\rangle = 0$ for all $i$,
eq.~(\ref{rellongtime}) is equal to
(within a factor of $\sqrt 2$) 
\begin{equation}\label{rellongtimeb}
  \max_{i = 1,\ldots,k} 
    \frac{|\langle\beta_i|P|\alpha\rangle|}{%
      \|P|\alpha\rangle\|\|\overline P|\alpha\rangle\|}\,. 
\end{equation}
The  cumulative frequency distribution for (\ref{rellongtimeb})
squared is calculated in~\cite{McElwaine:2} as
\begin{equation}
  P \left( \max_{i = 1,\ldots,k} 
    \frac{|\langle\beta_i|P|\alpha\rangle|^2}{%
      \|P|\alpha\rangle\|^2 \|\overline P|\alpha\rangle\|^2} 
    < \lambda \right ) =  
  \sum_m (-1)^m {k \choose m} (1-m\lambda)^{d-1} 1_{m\lambda<1},
\end{equation}
which approximately equals $[1-e^{-d \lambda}]^k$ when $dk^2\lambda^2
= o(1)$. This is the probability that a pair of projections acting on
one history in a set of $k+1$ consistent histories satisfies the
medium DHC with parameter $\epsilon = \sqrt\lambda$.  There are $n_p =
2^{\min(d_1-1,d_2)}-1$ distinct choices for the projections so the DHP
(to within a factor of $\sqrt2$) for extending $\alpha$ with any
Schmidt projection is
\begin{equation}\label{rellongtimec}
  \min_{j = 1,\ldots,n_p} \max_{i = 1,\ldots,k} 
    \frac{|\langle\beta_i|P_j|\alpha\rangle|}{%
      \|P_j|\alpha\rangle\|\|\overline P_j|\alpha\rangle\|}\,, 
\end{equation}
where $\{P_j, \overline P_j\}$ range over all $n_p$ binary partitions
of the basis Schmidt projections.  The distribution for this random variable
is hard to calculate but if we assume that the DHP's for each $\{P_j,
\overline P_j\}$ are independent the cumulative distribution function
is
\begin{equation}\label{rellongtimed}
  1-\{1-[1-e^{-d \lambda + o(1)}]^k\}^{n_p}\,.
\end{equation}
This assumption is obviously very approximate since the different
projections are all formed using the same basis. However, treating the
$\{P_j, \overline P_j\}$ as independent in (\ref{rellongtimec}) will be
a lower bound for the exact result and (\ref{rellongtimeb}) will be an
upper bound for the exact result.

Suppose now we wish to choose $\epsilon$ so that the probability of
making a projection at a large time is $p$, where $p$ is close to one. Then
from eq.~(\ref{rellongtimed})
\begin{equation}
  \epsilon^2 = -1/d\log\{1- [1-(1-p)^{1/n_p}]^{1/k}\}\,.
\end{equation}
For large $k$
\begin{equation}
 \epsilon^2 \approx -1/d\log\{-1/k\log[1-(1-p)^{1/n_p}]\} \approx
 1/d\log(k)\,.
\end{equation}
This calculation has involved a lot of assumptions and approximations,
but it should accurately reflect the behaviour for $d \gg k^2 \gg 1$.
The logarithmic dependence on $k$ is a generic feature of extreme
order statistics and so is the independence of the answer from the other
factors $p$ and $n_p$. The $1/\sqrt d$ dependence is also expected because
the mean value of the square of the inner product between two random
vectors in a $d$ dimensional space is $1/d$. 

A generic feature of asymptotic extreme order statistics such as the
previous calculation, is the slow rate of convergence and this
calculation is only expected to be accurate for very large $d$ and
$k$. For actual application to particular models the exact
distribution can be calculated by a computer program using Monte-Carlo
methods.

For each sample the program randomly picks a set of exactly consistent
histories, and then calculates the DHP for all $n_p$ combinations of
projections on one particular history. Ten thousand samples were
sufficient to produce smooth cumulative frequency distributions.
These can be inverted to calculate $\epsilon(p,k)$ as in
fig.~(\ref{fig:percentile}), where for example $\epsilon(.99,k)$ is
the solid curve.

The same arguments apply for absolute consistency as $t \to \infty$,
but since the consistency requirement is not normalised the expected
DHP values will be reduced by a factor of $1/\sqrt k$ --- the
average value for the length of a history when there are $k$
histories.

These calculations suggest that choosing $\epsilon = O(1/\sqrt d)$ is
most likely to produce histories with a complicated branching
structure and many non-trivial projections. In the next section we
discuss the results of simulations for all values of the parameters
and show that they agree with these theoretical calculations.

\section{Computer simulations}
\begin{figure}%
  \includegraphics{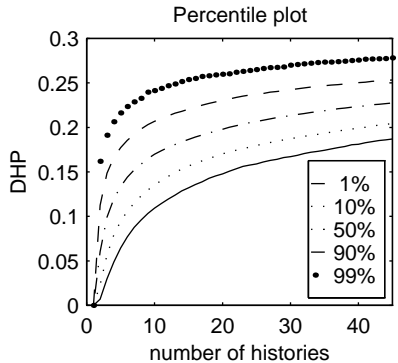}%
  \caption{estimating $\epsilon$}\label{fig:percentile}%
\end{figure}

The computer programs are explained and listed in
ref.~\cite{McElwaine:2}. The results described here were carried out
with a system of dimension $3$, with an environment of dimension $15$
and with either medium absolute consistency or medium relative
consistency (DHC). Fig.~(\ref{fig:percentile}) gives the probability
distribution for the DHP plotted via percentile curves as a function
of the number of histories in the long time limit.  For example, this
graph shows that with $\epsilon \geq 0.3 \approx 2/\sqrt d$
projections will almost certainly be consistent for any number of
histories, whereas for $\epsilon \leq 0.05 \approx.3/\sqrt d$
projections will probably only be consistent when there are two or
three histories. When there are twenty histories it shows that for
$98\%$ of the time the DHP will be between $0.15$ and $0.25$.
Fluctuations will only occasionally ($1\%$ of the time) lie below the
solid line and if $\epsilon$ has this value (for a particular number
of histories) then although projections will probably occur they will
occur as a result of large fluctuations from the mean. Therefore one
would expect the projections to occur at widely separated times and if
$\epsilon$ is changed only slightly the times generically to change
completely, and indeed computer simulations show this.

The simulations described here were run for ten thousand program steps
or until thirty histories had been generated. For a given set of
parameters, many simulations with different Hamiltonians and initial
states were carried out and were found generically to produce
qualitatively the same results, though only individual
simulations are described here.

\pairoffigures[tree]{ds}{example}{example} One way to look at the
results of a simulation is to look at the \emph{probability tree}
associated with the set of histories such as
fig.~(\ref{fig:example}a). The root node on the far left represent the
initial state, the terminal nodes represent the histories and the
other nodes represent intermediate path-projected states. Each node
has a probability and the lines linking the nodes have an associated
projection operator and projection time. The projections associated
with lines emanating to the right from the same node form a projective
decomposition and all occur at the same time. The scaling of the axis
and relative positions between the nodes is arbitrary, only the
topology is relevant. For example, in fig.~(\ref{fig:example}a) the
probabilities for the histories are $0.42$, $0.25$, $0.05$, $0.02$,
$0.02$, $0.01$, $0.08$ and $0.14$ --- the probabilities of the
terminal nodes.

Another useful interpretative aid is a graph of the \emph{consistency
  statistics} fig.~(\ref{fig:example}b). This graph shows the DHP for
the most consistent non-trivial extension. At times where no Schmidt
projections result in non-trivial histories no points are plotted,
though there are no such times in fig.~(\ref{fig:example}b).  The
program makes a projection when this value is $\epsilon$. The flat
line indicates $\epsilon$ and the crosses indicate when projections
have occurred --- in this case at times $0$, $1$, $11$, $12$, $36$ and
$65$ (approximately). A graph of the projection times will also be used
sometimes, for example fig.~(\ref{fig:releg2}b).

When any Schmidt eigenvalues are equal their eigenspaces becomes
degenerate and the corresponding Schmidt projections are not uniquely
defined. The reduced density matrix varies continuously in this model
and it will only be degenerate for a set of times of
measure zero so generically it is possible to define the Schmidt
states so that they are continuous functions of $t$ for all $t$.
This was not found to be necessary in the simulations.

\subsection{Results for relative consistency}
\subsubsection{Rank one initial density matrix} 
\pairoffigures{ds}{releg1}{relative consistency, $\epsilon =
  0.03 \approx 0.2/\sqrt{d}$} Fig.~(\ref{fig:releg1}) shows the
probability tree and minimum consistency statistics for a simulation
with $\epsilon=0.03 \approx 0.2/\sqrt d$ and a rank one initial
density matrix. As expected there is one almost immediate trivial
projection. Three more projections occur and then no more. From
fig.~(\ref{fig:percentile}) the probability of a projection with five
histories and $\epsilon = 0.03$ is less than 1\% so this result is as
expected. The simulation was run for longer than is shown in the
figure (until $t=100$) but no further projections occurred.

\pairoffigures{pjt}{releg2}{relative consistency, $\epsilon = 0.15
  \approx 1/\sqrt{d}$} \pairoffigures{pjt}{releg3}{relative
  consistency, $\epsilon = 0.16 \approx 1/\sqrt{d}$}
Fig.~(\ref{fig:releg2}) shows the probability tree and projection
times for a simulation with $\epsilon=0.15 \approx 1/\sqrt d$. Again
there is the initial trivial projection but no other trivial
projections occur. Projections then occurred at roughly equal equal
time intervals until there were fifteen histories.  The time between
projections then rapidly increased. This is in accord with
fig.~(\ref{fig:percentile}) as the probability for a projection with
fifteen histories and $\epsilon=0.15$ is around $5\%$. Projections after
this time only occur for large deviation away from the mean and
therefore occur extremely erratically. These later projections are
extremely unlikely to vary smoothly for a range of $\epsilon$. The
simulation was run until $t\approx100$ and no further projections
occurred. This simulation has produced an interesting set of histories
with a complicated branching structure.

The next pair of figures fig.~(\ref{fig:releg3}) shows the results of
a simulation with all the parameters unchanged except for $\epsilon$ which is
now $0.16$. The qualitative description is the same and the
first eight or so projections are similar. After that however the two sets of
histories are very different. This is the problem with the algorithm
applied to this model: interesting sets of histories are produced, but they
change dramatically for small changes in $\epsilon$.

\begin{figure}%
  \includegraphics{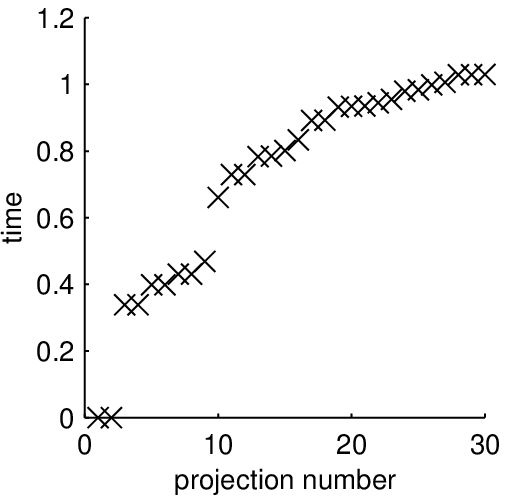}%
  \caption{relative consistency, $\epsilon=0.3\approx 2/\sqrt d$}
  \label{fig:releg4}
\end{figure}

From fig.~(\ref{fig:percentile}) choosing $\epsilon=0.3\approx 2\sqrt
d$ looks large enough so that projections will always be made before
the background level is reached. The theoretical analysis also
suggests that for such a large value of $\epsilon$ some repeated
projections will occur. Indeed fig.~(\ref{fig:releg4}) demonstrates
that nine repeated projections occurred each giving a history of
probability $\delta$. 

\onefigure[ds]{releg5}{relative consistency, $\epsilon$ chosen at
  $50\%$} An interesting alternative is to choose $\epsilon$ as a
percentile from fig.~(\ref{fig:percentile}), that is $\epsilon(k)
\approx \epsilon(e) \log k$ where $k$ is the number of histories.
Fig.~(\ref{fig:releg5}) demonstrates the consistency statistics for a
run with $\epsilon(k)$ chosen at the $50\%$ level. All of the
probabilities except for the initial projection were non-trivial.
Rather than the projections being made in regimes where the DHP
fluctuates about its mean value most of the projections have been made
at times when the DHP is monotonically decreasing, so that the
histories are much more likely to vary continuously with $\epsilon$.
Two other advantages of choosing $\epsilon$ this way are that larger
sets of histories are produced, and if an algorithm is designed to
produce a set of histories of a certain size choosing $\epsilon$ in
this way will produce a more consistent set than choosing $\epsilon$
to be constant.  However, though the results are more stable (when the
percentile is changed) than for constant $\epsilon$, results from
simulations show that they still change too much to single out a
definite set of histories.

By looking at the consistency statistics the problem is easy to
understand. Since a projection is made at the earliest possible time
generically once it has been made the DHP jumps up as the most
consistent projection has occurred. The consistency level then falls.
While it is decreasing monotonically any change in $\epsilon$ will
produce a continuous change in the time of the next projection.
However, if $\epsilon$ is too far below its mean level the projection
times will vary discontinuously, and all the projections afterwards
will generically be completely different. Since the mean level of the
DHP depends on the number of histories strongly either $\epsilon$ must
be chosen sufficiently large so as to be above this or it must be
chosen so as to increase with the number of histories. This is
demonstrated by the first few projections as shown in
fig.~(\ref{fig:releg7}) --- a close-up  of fig.~(\ref{fig:releg5}b)
would also show this.

\subsubsection{Other initial conditions}
\pairoffigures{ds}{releg7}{relative consistency, $\delta = 0.02$,
  $\epsilon = 0.05$} 
Simulations with a rank 2 initial state, and all
other parameters remaining the same, produce the same results except
that each of the initial projections is repeated producing two trivial
histories with probability $\delta$.  We can choose $\delta$ according
to eq.~(\ref{preventinitial:rel}) to try to prevent these projections,
that is choose $\delta = O(\epsilon^2)$. Since many of the histories
we expect to generate will have probabilities smaller than this it is
sensible to use a relative non-triviality criterion\footnote{Using a
  relative non-triviality criterion earlier does not qualitatively
  change the results except that the initial trivial history would
  also have been extended --- the results would have been
  qualitatively the same as for rank two initial reduced density
  matrices.}.  The analysis leading to eq.~(\ref{preventinitial:rel})
is only accurate to first order in $t$, therefor
eq.~(\ref{preventinitial:rel}) is only valid when $\delta$ is
sufficiently small. If $\delta$ is too large the consistency level of
a reprojection will start decreasing and when a reprojection
eventually becomes non-trivial it will be consistent. For the example
discussed there were no values of $\delta$ with $\epsilon=0.15$ that
prevent an initial trivial reprojection. Fig.~(\ref{fig:releg7})
demonstrates the start of a simulation with $\epsilon=0.05$ and
$\delta = 0.02 = 8\epsilon^2$. The graph of the consistency statistics
shows the initial projection at $t=0$ and then that there are no
non-trivial extensions until $t\approx 0.08$ by which time the
projection is not consistent. A non-trivial projection is made at
$t\approx 0.32$ and the algorithm then continues as before, with the
trivial projection avoided. Because the projection has not occurred
with probability $\delta$ there is a range of values for $\delta$ that
do not affect the resulting histories --- they are independent of
$\delta$. The first three projection times will obviously vary
continuously for a small range of $\epsilon$. The other projections
that occurred in this simulation all occurred at much more separated
times in a regime where the consistency level was not decreasing
monotonically.  If $\epsilon$ is chosen according to the percentile
distribution in fig.~(\ref{fig:percentile}) --- that is $\epsilon(k)
\approx \epsilon(e) \log k$ where $k$ is the number of histories ---
$\epsilon$ will be small enough initially to prevent any trivial
projections (with an initial density matrix of rank greater than one)
and will allow a full set of histories to be built up at later times.

Simulations with a full rank initial density matrix and
$\epsilon=0.15$ result in a trivial repeated projection for each
initial history. If $\epsilon$ is smaller ($\leq 0.07$) no trivial
reprojections occur. If two initial projections are made uncorrelated
with the Schmidt states then further (Schmidt) projections at $t=0$
will not generically be trivial or consistent. In both
cases, after the initial projections and possible reprojections the
qualitative behaviour is the same as the rank one case. 

\subsection{Results for absolute consistency}
To produce interesting sets of histories from an absolute consistency
criterion three effects need to be balanced against each other.  If
$\epsilon$ is too small the most likely projections will be those that
produce very small probability histories. If $\delta/\epsilon$ is too
small the likelihood of repeated projections (hence trivial histories)
will be high. If $\delta$ is too large the non-triviality criterion
will dominate the
algorithm and only probability $\delta$ (trivial) histories will be
produced. Only an absolute parameterised non-triviality criterion is
considered since a relative criterion will clearly produce almost
nothing except infinitesimal histories. The following results
demonstrate these effects. 

\subsubsection{Rank one initial density matrix}
\begin{figure}%
  \includegraphics{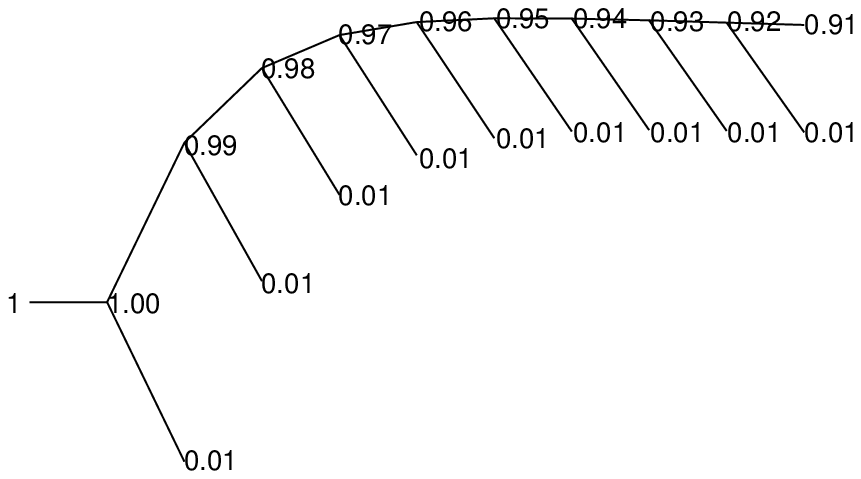}%
  \caption{absolute consistency, $\delta=0.01$, $\epsilon=0.1$}
  \label{fig:abseg1}
\end{figure}
For example if $\delta < \epsilon$ and the initial reduced density
matrix has rank one the algorithm will generically produce
$\lfloor1/\delta\rfloor$ trivial histories. This is a particularly
simple case of the analysis that suggests that if $\delta/\epsilon <
O(1)$ repeated projections are probable. 
Fig.~(\ref{fig:abseg1}) shows an example of this from a computer
simulation with $\epsilon =.1$ and $\delta=0.01$. Only the first ten
projections are shown. This behaviour remains the same in the limit as
$\epsilon \to 0, \delta \to 0, \delta\leq\epsilon$.

\onefigure{abseg12}{absolute consistency, $\delta=0.01$,
  $\epsilon=0.02$} As the ratio $\delta/\epsilon$ increases and
becomes $O(1)$ the nature of the set of histories changes.
Occasionally when a reprojection becomes non-trivial it will no longer
be consistent and a reprojection will not occur. A significant time
may elapse before the next projection is made which will result in a
non-trivial projection, which will then be followed by more trivial
repeated projections.  This is demonstrated in
fig.~(\ref{fig:abseg12}) where $\epsilon = 0.02$ and $\delta=0.01$.
Though this is an interesting set of histories this range of parameter
values does not give a theory with predictive power since simulations
show that the results vary enormously for small changes in $\epsilon$
and $\delta$.

\pairoffigures{ds}{abseg4}{absolute consistency, $\delta=0.01$,
  $\epsilon=0.001$} As $\delta/\epsilon$ increases past $1$ the number
of histories made with probability $\delta$ decreases to just the
initial projection. Fig.~(\ref{fig:abseg4}) shows that this range of
parameter values produces interesting histories but the projections
are occuring at times when the consistency level is fluctuating
randomly about the mean and so will be unstable to small changes in
$\epsilon$.

\subsubsection{Other initial conditions}
Choosing larger rank initial reduced density matrices or initial
projections does not qualitatively change the analysis. The only
difference is that for $\delta > \epsilon$ and $\epsilon$ sufficiently
small no trivial projections will be made.
 
\section{Conclusions}
The algorithm produces sets of histories with a complicated branching
structure and with many non-trivial projections for a range of
parameter values. Algorithms using the DHC produce results that are
essentially the same for a wide range of $\delta$ (the non-triviality
parameter) including the limit $\delta \to 0$. However, the algorithm
does not make useful predictions when applied to this model since the
results vary erratically with $\epsilon$ and there is no special
choice of $\epsilon$ singled out. Choosing $\epsilon$ as a function of
the number of histories according to fig.~(\ref{fig:percentile}) produces
the least unstable sets of histories and the largest sets of
non-trivial histories, but even in this case the algorithm does not
single out a definite set.

The algorithm is less effective when used with an absolute consistency
criterion: in this case the predictions of the algorithm also vary
erratically with $\delta$ and the resulting sets of histories include
fewer non-trivial histories. 

The results of the simulations agree well with the theoretical
analysis of section~(\ref{sec:rand:analysis}) and demonstrate features
of the algorithm that will also apply to other models --- such as the
analysis of repeated projections. They also demonstrate some of the
difficulties that an algorithm must overcome. These problems can be
related to the discussion of recoherence in
refs.~\cite{McElwaine:Kent}.  The algorithm will only produce stable
results (with respect to $\epsilon$) if the projections occur when the
off-diagonal terms of the decoherence matrix are monotonically
decreasing. This behaviour is only likely in a system like this for
times small compared to the recurrence time of the system and when the
number of histories is small compared to the size of the environment
Hilbert space. The results of the model do show stability for the
first few projections and if much larger spaces were used this
behaviour would be expected for a larger number of histories. In
particular as the size of the environment goes to infinity it is
plausible that the algorithm applied to this model will produce a
large, stable, non-trivial set of histories.

In retrospect it was ambitious to hope that an algorithm applied to
this model would produce large sets of stable histories. A random
model like this, with such a small environment, will generically only
decohere a few histories (see
also~\cite{Albrecht:decoherence,Albrecht:collapsing}.)

\appendix
\section*{The Gaussian Unitary Ensemble}\label{app:gue}
In the GUE the matrix elements are chosen according to the
distribution
\begin{eqnarray*}
  p(A) & = & \frac{2^{n/2}}{[(2\pi)^{1/2}\sigma]^{n^2}}
  \exp\left\{-\frac{\mbox{Tr}(A^2)}{4\sigma^2}\right\}\\ & = &
  \frac{2^{n/2}}{[(2\pi)^{1/2}\sigma]^{n^2}} \prod_{n\geq j,k \geq
  1}\exp\left\{-\frac{A_{jk}A_{kj}}{4\sigma^2}\right\}\\ & = &
  \frac{2^{n/2}}{[(2\pi)^{1/2}\sigma]^{n^2}} \prod_{n\geq j
  \geq=1}\exp\left\{-\frac{X_{jj}^2}{4\sigma^2}\right\} \prod_{n\geq k
  \geq j \geq 1} \exp\left\{-\frac{X_{jk}^2}{2\sigma^2}\right\}
  \prod_{n\geq k \geq j \geq
  1}\exp\left\{-\frac{Y_{jk}^2}{2\sigma^2}\right\}
\end{eqnarray*}
where $ A_{jk} = X_{jk} + i Y_{jk}$, $X_{jk} = X_{kj}$ and $Y_{jk} =
-Y_{kj}$. Therfore all the elements are independently, normally
distributed, the diagonal with variance $2\sigma$ and the real and
imaginary off-diagonal with variance $\sigma$. Some expectations for a
normal variable with variance $\sigma$ are
\begin{eqnarray*}
  \mbox{E}[|X|^{n}] & = &
  \frac{2^{n/2}\sigma^n\Gamma(\frac{p+1}{2})}{\sqrt{\pi}}
\end{eqnarray*}
and in particular $\mbox{E}[|X|] = \sqrt{(2/\pi)}\sigma$,
$\mbox{E}[X^2] = \sigma^2$ and $\mbox{E}[X^4] = 3\sigma^2$.  Since
$X_{ij}$ is independent of $X_{kl}$ unless $i=k$ and $j=l$, or $i=l$
and $j=k$
\begin{eqnarray*}
  \mbox{E}[X_{ij}X_{kl}] & = & \sigma^2(\delta_{il}\delta_{jk} +
  \delta_{ik}\delta_{jl}),\\ \mbox{E}[Y_{ij}Y_{kl}] & = &
  \sigma^2(\delta_{ik}\delta_{jl} - \delta_{il}\delta_{jk}).
\end{eqnarray*}
Therefore, for the elements of $A$
\begin{eqnarray*}
  \mbox{E}[A_{ij}] & = & 0, \\ \mbox{E}[A_{ij}A_{kl}] & = &
  2\sigma^2\delta_{il}\delta_{jk}, \\
  \mbox{E}[A_{ij}A_{kl}A_{mn}A_{op}] & = & 4\sigma^4
  (\delta_{il}\delta_{jk}\delta_{mp}\delta_{no} +
  \delta_{in}\delta_{jm}\delta_{kp}\delta_{lo} +
  \delta_{ip}\delta_{jo}\delta_{kn}\delta_{lm}).
\end{eqnarray*}
Applying these results to vectors and projection operators,
\begin{eqnarray}\label{expectgue}
  \mbox{E}[{\bf n}^\dagger A{\bf m}] & = & 0, \\\nonumber
  \mbox{E}[|{\bf n}^\dagger A{\bf m}|^2] & = & 2\sigma^2 |{\bf
  n}|^2|{\bf m}|^2,\\\nonumber \mbox{E}[{\bf n}^\dagger APA{\bf m}] &
  = & 2d\sigma^2{\bf n}^\dagger{\bf m},\\\nonumber \mbox{E}[|{\bf
  n}^\dagger APA{\bf m}|^2] & = & 4\sigma^4 [ d^2|{\bf n}^\dagger{\bf
  m}|^2 + ({\bf n}^\dagger P{\bf n}) ({\bf m}^\dagger P{\bf m}) + d
  |{\bf n}|^2|{\bf m}|^2],
\end{eqnarray}
where $d$ is the rank of $P$. For the real part or the imaginary part
just take half of the above since $|z|^2 = [\mbox{Re}(z)]^2 +
[\mbox{Im}(z)]^2$.


\end{document}